\documentclass[a4paper, 12pt]{article}
\usepackage{graphicx}
\newcommand{\la}{\langle}
\newcommand{\ra}{\rangle}

\renewcommand{\d}{\partial}
\newcommand{\beq}{\begin{eqnarray}}
\newcommand{\eeq}{\end{eqnarray}}
\newcommand{\sbeq}{\begin{subeqnarray}}
\newcommand{\seeq}{\end{subeqnarray}}

\newcommand{\delslash}{\gamma\cdot\d}

\newcommand{\Sigpi}{\Sigma_{\pi N}}

\newcommand{\btem}{\bibitem}

\newcommand{\pipi}{{$\pi$-$\pi$}}
\newcommand{\chis}{chiral symmetry }

\newcommand{\THK}{T. Hatsuda\ and \ T. Kunihiro\ }
\newcommand{\HK}{T. Hatsuda\ and \ T. Kunihiro\ }
\newcommand{\KH}{ T. Kunihiro and T. Hatsuda\,  }
\newcommand{\PR}{Phys. Rev. }
\newcommand{\PL}{Phys. Lett. {\bf B}}

\title{
 The Sigma Meson and Chiral Transition in Hot and Dense Matter
}

\author{Teiji Kunihiro\\
Yukawa Institute for Theoretical Physics,\\
 Kyoto University, Sakyo-ku, Kyoto 606-8502, Japan}

\begin{document}

\maketitle

\begin{abstract}
It is pointed out that the hadron spectroscopy  should be 
a study of the structure of the QCD vacuum, low-energy elementary
excitations on top of which are hadrons.
Concentrating on the dynamical breaking of the
 chiral symmetry in the QCD vacuum,
we emphasize the importance to clarify what is going on with mesons in 
the $I=J=0$-channel, i.e., the sigma meson channel, because it is 
connected to the quantum fluctuations of the chiral order parameter.
After summarizing the significance of the sigma meson in  QCD and
low-energy hadron phenomenology,
we give a review on  some theoretical and experimental effort to try to 
reveal the possible restoration of chiral
symmetry in hot and dense nuclear matter including heavy nuclei. 
\end{abstract}

\setcounter{section}{0}
\section{Introduction}

A tricky point in  the hadron spectroscopy is that QCD, 
the fundamental theory of the hadron world,
 is not written in terms of  hadron fields but 
 in terms of  quark- and gluon-fields from which
 hadrons are composed:
The quarks and gluons are colored objects which 
can not exist in the asymptotic state, 
and  low-lying elementary excitations
 on top of the non-perturbative QCD vacuum are composite
and  colorless, which we call hadrons.
Furthermore, symmetries possessed by the QCD Lagrangian,
such as the chiral SU(3)$_L\times$SU(3)$_R$ symmetry
 in the massless limit of quarks and the color gauge symmetry
are not manifest in our every-day world.
This complication of the problem is due to the fact that 
the true QCD vacuum is completely different from the 
perturbative one and is actually realized through the phase 
transitions, i.e.,
 the confinement-deconfinement and the chiral transitions.
The notion of such a complicated vacuum structure, 
i.e., 
 the collective nature of the vacuum and the elementary particles
was first introduced by Nambu\cite{nambu}, in analogy with the
physics of superconductivity\cite{nambu2}.
The nonperturbative nature and the realization of the
true QCD vacuum through the phase transitions are being 
confirmed by the lattice simulations\cite{karsch}.
One may  notice that the so called $U_A(1)$ anomaly\cite{anomaly}
 also characterizes the non-perturbative QCD vacuum.
Several rules extracted from the hadron 
phenomenology such as the vector-meson
dominance (VMD)\cite{vmd,bando} and the Okubo-Zweig-Iizuka (OZI)
rule\cite{ozi} might be 
also related with some fundamental properties of the QCD vacuum.
Thus the  hadron spectroscopy can not be failed to be  
a  study of the nature of QCD vacuum including 
its symmetry properties. In other words, the physics of the hadron 
spectroscopy is a combination of the condensed matter physics of the 
QCD vacuum\cite{hk94,rw00}
 and the atomic physics as played with the constituent
quark-gluon model where the vacuum structure is taken for 
granted\cite{consti}.

An interesting observation  is then that  hadrons as elementary excitations
 on top of the QCD vacuum may change their properties 
in association with a change or phase transition of the QCD 
vacuum.
What hadrons do change their properties sharply, and 
 how do they  in hot and/or dense medium?
One should also ask how  they are detected in 
experiment\cite{pisarski,hk84,hk85,miyamura,br91};
 see also the reviews
\cite{hk94,br96}.
For instance, in Table 6.1 in \cite{hk94}, list up are 
interesting observables and their expected 
behavior in relation with the chiral transition,  
possible restoration of the $U_A(1)$-symmetry and 
precritical deconfinement
at finite temperature and/or density.
In association with (partial) restoration of chiral symmetry, 
the mass of the $\sigma$ meson\cite{hk85,bernard87}
 is expected to decrease.  
 Some people\cite{pisarski,br91,hkl}
 expect that the vector mesons $\rho, \omega$
 and $\phi$ also show a decrease of their masses 
in association with the chiral restoration.
 The $U_A(1)$ anomaly, which is responsible for 
 lifting the $\eta '$ meson mass as high as about 1 GeV and make the 
 $\eta' (\eta)$ almost flavor singlet (octet), may be cured at high 
temperature, which may manifest itself as the decrease of the mass 
$m_{\eta'}$\cite{pw84,kuni89}
 for example.  The deconfinement may affect the 
properties of heavy-quark systems such as 
J$/\psi$\cite{miyamura,matsui} than in light hadrons.

In the present talk, I  will focus on the
chiral transition in hot and dense hadronic matter
 and discuss the significance of the
 scalar and isoscalar meson, the sigma meson and the strength function
 in its channel in the hadronic medium including heavy nuclei.

\section{The sigma meson}
\setcounter{section}{2}
The order parameter of the chiral transition
is the quark condensate $\la \bar {q}q\ra\sim\sigma$.
There arise two kinds of quantum fluctuations
of the order parameter, the modulus and phase fluctuations.
The pion corresponds to the latter fluctuations.
The particle corresponding to the former fluctuation
is a scalar-isoscalar meson, which is traditionally called
the $\sigma$ meson.
One may notice that the way of the 
appearance of the $\sigma$ meson is
analogous to that of the Higgs particle,  
which comes to exist through 
the dynamical breaking of the gauge symmetry in the 
standard model while the corresponding NG boson is 
absorbed into the longitudinal component of the gauge fields.
As one can now see, 
{\em the existence of the $\sigma$ meson is logically 
related to the fundamental property of the QCD vacuum
 in which  the chiral symmetry is spontaneously broken.}
Therefore searching for such a particle in experiment is 
{\em not eccentric but as naturally motivated as searching for 
glue balls in QCD and the Higgs particle in the standard model.}  
If such a particle or the like could not be identified, one must
consider possible dynamical origins to hinder them from appearing.


Here are a short summary of
 the significance of the sigma meson in hadron physics:
\begin{enumerate}
\item
 The existence of the $\sigma$ meson as 
{\em the quantum fluctuation of the
 order parameter} of the chiral transition
 accounts for various  phenomena in hadron physics
which otherwise remain mysterious\cite{elias,hk94,supple}.
\item
There have been accumulation of 
 experimental evidence of a low-mass
 pole in the 
$\sigma$ channel in the pi-pi scattering matrix\cite{pipiyitp,pipi}. 
It should be emphasized that for obtaining this result,
it is essential to respect chiral symmetry, 
analyticity and crossing symmetry even in an approximate way as 
in the $N/D$ method\cite{igi}:
The great achievement of the chiral perturbation theory\cite{chipert} is
indispensable for set the precise boundary condition for the
 scattering matrix in the low-energy region.
\item
It is well known that such a scalar meson with the 
mass range 500 to 700 MeV is 
responsible for the intermediate range attraction in
the nuclear force.\cite{sawada}

\item
The correlation in the scalar channel as summarized 
by such the sigma  meson may account 
for the enhancement of the $\Delta I=1/2$ processes in 
 K$^{0} \rightarrow \pi^{+}\pi^{-}$ or 
 $\pi^{0}\pi^{0}$ \cite{morozumi}.
In fact,  the final state interaction for the emitted two pions
may include the $\sigma$ pole, then
 the matrix element for of the scalar 
operator $Q_6\sim \bar{q}_Rq_L\bar{q}_Lq_R$ is sown to be
enhanced dramatically.

\item

The collective excitation in the scalar channel
 as described as   the $\sigma$ meson is essential 
\cite{kuni90} in reproducing the empirical value of the $\pi$-N 
sigma term \cite{JK}
 $\Sigpi =\hat {m}\la \bar{u} u + \bar {d} d\ra$:
Empirically, it is known that
 $\Sigma_{\pi N}\sim$ 40-50 MeV, while the naive quark model
only gives as small as  $\sim 15$ MeV.
The basic quantities here are the 
quark contents of baryons $\la B\vert \bar {q}_i q_i\vert B\ra \equiv
 \la \bar{q}_iq_i\ra_B$ ($i= u, d, s, ...$).  Actually, it is more 
adequate to call them the scalar charge of the hadron. 
The point is such a scalar charge of the nucleon is 
enhanced with the existence of the sigma meson pole;
 such an enhancement of charges by collective modes are well
known in nuclear physics.
\end{enumerate}

\setcounter{equation}{0}
\setcounter{section}{3}
\section{Partial chiral restoration and the 
$\sigma$ meson in  hadronic matter}

Although the recent phase shift analyses \cite{pipi} 
of the \pipi scattering 
and the identification of the pole 
in the $I=J=0$ channel as mentioned in 2 above
is a great development in this field, 
one must say that it is still obscure whether
the pole really corresponds to the quantum fluctuation of the 
chiral order parameter, i.e., our $\sigma$.
If one were to be able to change the environment freely and trace the
possible change of the pole position,
the nature of the particle corresponding to the pole
can be revealed: A change of the vacuum or the equilibrium
 state leading to the phase transition will make 
the mode coupled to the order parameter change.
Actually this is the usual strategy in the many-body
physics\cite{soft} to reveal the nature of elementary 
excitations.
Conversely, an observation of the change of the elementary
modes as well as that of other thermodynamic quantities 
tells us the change of the state of the matter.

Effective theories of QCD\cite{hk85,bernard87} show that 
 the sigma meson mass $m_{\sigma}$ decreases in association with 
the chiral  restoration in hot and/or dense 
medium, while the pion mass keeps its value in free space as long
 as the system is in the Nambu-Goldstone phase.
The simulations  on the lattice QCD also  show 
a decrease of the {\em screening mass} in the sigma meson channel;
see for instance, \cite{karsch,scrlattice}. The screening mass which 
describes the damping of the correlation function in the space
 direction is not the dynamical 
 mass which is given through the time correlation of the relevant
operators. Nevertheless, it is remarkable that the lattice result is 
not in contradiction with those in the effective theories.

 Then the width of 
the $\sigma$ is also expected to decrease  due to the
 depletion of the phase space for the decay 
$\sigma \rightarrow 2\pi$\cite{hk87}.
Thus one can  expect a  chance to see the $\sigma$ meson as a sharp 
resonance at high temperature and/or density.
  
Some years ago, the present author
proposed several nuclear experiments including
 one using electro-magnetic probes 
to produce the $\sigma$ meson in nuclei, thereby 
have a clearer evidence of  the  existence of 
the $\sigma$ meson and also explore the possible 
restoration of chiral  symmetry in the nuclear
 medium\cite{supple,tit}:
As is well known, there arises a scalar-vector mixing 
 in nuclear matter at finite density\cite{weldon}.
To make a veto for the two pions from the rho meson, the produced
pions should be neutral ones which may be detected through
four $\gamma$'s.

When a hadron is put in a hadronic medium, the hadron might 
dissociate into complicated excitations to loose its identity.
Then the most informative quantity is the response function 
or spectral function in the hadron channel of the system.
If the coupling of the hadron  with the environment is relatively 
small, then there may remain a peak with a small width in the 
spectral function corresponding to the hadron.
Such a peak is to be identified with
 an elementary excitation or a quasi particle.
Then how will the decrease of $m_{\sigma}$ in the nuclear medium
affect  the spectral function.

It has been shown by using linear chiral models 
that an enhancement in the spectral function in the
 $\sigma$ channel occurs just above the
two-pion threshold  along with the decrease of 
$m_{\sigma}$\cite{CH98}. Subsequently, 
it has been shown \cite{HKS} that the spectral enhancement
near the $2m_{\pi}$ threshold takes place 
in association with  partial restoration of \chis  
at finite baryon density.

Hatsuda et al\cite{HKS}
started from  the following linear sigma model;
\beq
\label{model-l}
{\cal L}  =   {1 \over 4} {\rm Tr} [\partial M \partial M^{\dagger}
 - \mu^2 M M^{\dagger} - {2 \lambda \over 4! } (M M^{\dagger})^2
 -  h (M+M^{\dagger}) ]
  + \bar{\psi} ( i \delslash - g M_5 ) \psi
  + \cdot \cdot \cdot  ,
\eeq
where  $M = \sigma + i \vec{\tau}\cdot \vec{\pi}$,
 $M_5 = \sigma + i \gamma_5 \vec{\tau}\cdot \vec{\pi}$,
 $\psi$ is the nucleon field, and
 Tr is for the flavor index.
Consider the propagator 
 of the $\sigma$-meson at rest in the medium :
$D^{-1}_{\sigma} (\omega)= \omega^2 - m_{\sigma}^2 - $
$\Sigma_{\sigma}(\omega;\rho)$,
where $m_{\sigma}$ is the mass of $\sigma$ in the tree-level, and
$\Sigma_{\sigma}(\omega;\rho)$ is 
the loop corrections
in the vacuum as well as in the medium.
 The corresponding spectral function is given by 
$\rho_{\sigma}(\omega) = - \pi^{-1} {\rm Im} D_{\sigma}(\omega)$.
One can show that 
\beq
{\rm Im} \Sigma_{\sigma}
\propto \theta(\omega - 2 m_{\pi})
 \sqrt{1 - {4m_{\pi}^2 \over \omega^2}}
\eeq
near the two-pion threshold  in the one-loop order.
On the other hand, partial restoration of \chis 
implies that $m_{\sigma}^*$ 
 defined by
${\rm Re}D_{\sigma}^{-1}(\omega = m_{\sigma}^*)=0$
  approaches to $ m_{\pi}$.  Therefore,
 there exists a density $\rho_c$ at which 
 ${\rm Re} D_{\sigma}^{-1}(\omega = 2m_{\pi})$
 vanishes even before the complete restoration
 of \chis where $\sigma$-$\pi$
 degeneracy is realized,
 namely 
${\rm Re} D_{\sigma}^{-1} (\omega = 2 m_{\pi}) =
 [\omega^2 - m_{ \sigma}^2 -
 {\rm Re} \Sigma_{\sigma} ]_{\omega = 2 m_{\pi}} = 0$.
At this point, the spectral function is solely 
given in terms of the
 imaginary part of the self-energy;
\beq
\rho_{\sigma} (\omega \simeq  2 m_{\pi}) 
 =  - {1 \over \pi \ {\rm Im}\Sigma_{\sigma} }
 \propto {\theta(\omega - 2 m_{\pi}) 
 \over \sqrt{1-{4m_{\pi}^2 \over \omega^2}}},
\eeq
which clearly shows the near-threshold enhancement of
the spectral function.
 This is a general phenomenon correlated with the 
 partial restoration of \chis.

\begin{center}
\begin{figure}
\scalebox{0.35}{%
\includegraphics{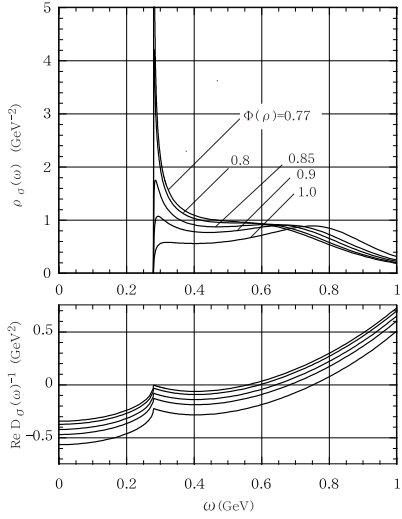}
}
\label{fig.1}
\caption{The spectral function $\rho_\sigma(\omega)$ (the upper panel)
and ${\rm Re} D_{\sigma}^{-1}(\omega)$  (the lower panel)
 calculated with a linear sigma model.
$\Phi(\rho)\equiv \langle \sigma \rangle/ \sigma_0$ measures
 the rate of the partial restoration of the chiral symmetry at the
baryonic density $\rho$. }
\end{figure}
\end{center}
In \cite{HKS},
 the effect of the meson-loop as well as
 the baryon density  was treated as a perturbation 
 to the vacuum quantities. Therefore, our loop-expansion  
  is valid only at relatively low
 densities.
When we parameterize the chiral condensate in nuclear matter
 $\langle \sigma \rangle$ as
\beq
\langle \sigma \rangle \equiv  \sigma_0 \ \Phi(\rho),
\eeq
 one may take the linear density approximation for 
small density;
$\Phi(\rho) = 1 - C \rho / \rho_0 $
with
 $C = (g_{\rm s} /\sigma_0 m_{\sigma}^2) \rho_0$.

%
%

The spectral function $\rho_\sigma(\omega)$ together with 
${\rm Re} D_{\sigma}^{-1}(\omega)$  
 calculated with a linear sigma model are shown 
  in Fig.1: The characteristic enhancements of the spectral
 function is seen just above the 2$m_{\pi}$.
It is also to be noted that even before
the $\sigma$-meson mass $m_{\sigma}^*$ and $m_{\pi}$ in the medium 
are degenerate,i.e., the chiral-restoring point, 
 a large enhancement
 of the spectral function near the $2m_{\pi}$ is seen.

Is the near-threshold enhancement obtained above
 specific to the linear representation of the
 chiral symmetry, where the $\sigma$ degree of freedom is explicit
 as in (1).
Jido et al\cite{jhk}
showed that the nonlinear realization of the chiral 
symmetry can also
give rise to the near 2$m_{\pi}$ enhancement of the
spectral function in nuclear medium as shown in Fig.2.
\begin{center}
\begin{figure}
\scalebox{0.6}{%
\includegraphics{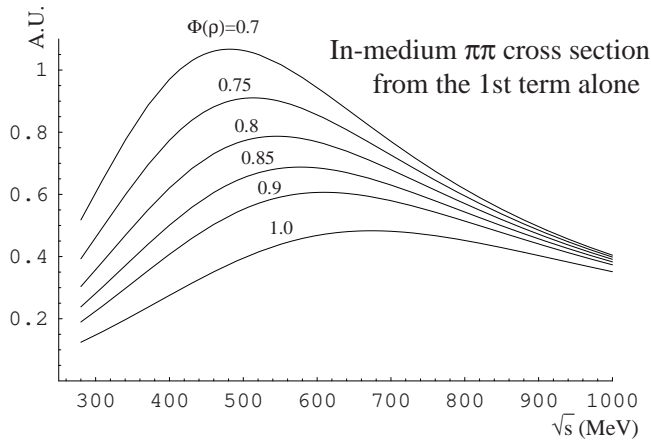}
}
\caption{
In-medium $\pi\pi$ cross section in the $I=J=0$ channel
 in the heavy $S$ limit where $m_{\sigma}^*$ is taken to be infinity.
    The cross section is shown in the arbitrary unit (A.U.).
 }
\label{fig2}
\end{figure}
\end{center}
They begin the discussion  with the polar parameterization of the chiral field,
 $M = \sigma + i \vec{\tau} \cdot \vec{\pi}
  = (\la \sigma \ra + S) U $ with $U = \exp (i \vec{\tau}
 \cdot \vec{\phi} /f^{*}_{\pi})$. 
Here $f^{*}_{\pi}$ is a would-be ``in-medium pion decay constant''.
\beq
\label{model-nl}
{\cal L}& =&   {1 \over 2} [(\partial S)^2 - m_{\sigma}^{*2} S^2]
  - {\lambda \la \sigma \ra \over 6} S^3 - {\lambda \over 4!} S^4 
  +  {(\la \sigma \ra +S)^2 \over 4} {\rm Tr}
 [\partial U \partial U^{\dagger}] + { \la \sigma \ra + S \over 4}\  h \
 {\rm Tr}[U^{\dagger}+U] \nonumber \\
& + & {\cal L}_{\pi N}^{(1)} - g S \bar{N} N \ ,
\eeq
with
${\cal L}_{\pi N}^{(1)} =
\bar{N}(i \delslash + i v \hspace{-6pt}/
 + i  a \hspace{-6pt}/   \gamma_{5}  - m_{N}^* ) N$
and $(v_{\mu},a_{\mu}) = (\xi \partial_{\mu} \xi^{\dagger}
 \pm \xi^{\dagger} \partial_\mu \xi)/2$, and
 $m_N^* = g \la \sigma \ra$.
In this representation, the in-medium $\pi\pi$ amplitude in the tree
level
reads
 \beq
\label{scatt2}
A(s) =  {s- m^2_\pi \over \la \sigma \ra^2}
       - {(s - m_\pi^2)^2 \over
 \la \sigma \ra^2} {1 \over s - m_{\sigma}^{*2}}  .
\eeq
The first term in (\ref{scatt2}) comes from the
 contact 4$\pi$ coupling generated by  the
 expansion of the second line in (\ref{model-nl}) with
 the coefficient proportional to
 $1 / {\la \sigma \ra}^2$.
 On the other hand,
 the second term in (\ref{scatt2}) is from the contribution of the
  scalar meson $S$ in the $s$-channel.
Fig.2 shows a unitarized in-medium $\pi\pi$ cross section only
 with the first term in (\ref{scatt2}), i.e., what is given by the 
non-linear realization. One sees a clear enhancement
  of the cross
section near the threshold or a softening as chiral symmetry is restored.
Although there is no explicit $\sigma$-degrees of freedom in this 
heavy $\sigma$ approximation, 
 there arises a decrease of the
 pion decay constant $f_{\pi}^{\ast}$ in nuclear medium.
This  is due to a new vertex, i.e.,
4$\pi$N-N
 vertex absent in the free space; see Fig. 3.
The vertex is responsible
for the reduction of  $f_{\pi}^{\ast}$ and hence for 
the spectral enhancement.


\begin{center}
\begin{figure}
\scalebox{0.6}{%
\includegraphics{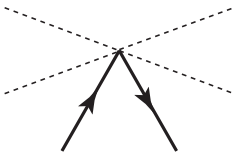}
}
\caption{
The new 4$\pi$-$N$-$N$ vertex generated
 in the nonlinear realization.
 The solid line with arrow and the dashed line represent
 the nucleon and pion, respectively.}
\label{fig.3}
\end{figure}
\end{center}

\section{Possible experimental evidence}
Interestingly enough,
CHAOS collaboration  \cite{chaos} had measured the 
$\pi^{+}\pi^{\pm}$
invariant mass distribution $M^A_{\pi^{+}\pi^{\pm}}$ in the
 reaction $A(\pi^+, \pi^{+}\pi^{\pm})X$ with the 
 mass number $A$ ranging
 from 2 to 208: They observed that
the   yield for  $M^A_{\pi^{+}\pi^{-}}$ 
 near the 2$m_{\pi}$ threshold is close to zero 
for $A=2$, but increases dramatically with increasing $A$. They
identified that the $\pi^{+}\pi^{-}$ pairs in this range of
 $M^A_{\pi^{+}\pi^{-}}$ is in the $I=J=0$ state.
The $A$ dependence of the 
 the invariant mass distribution presented in \cite{chaos} 
 near 2$m_{\pi}$ threshold has a close
 resemblance to our model calculation in Fig.1, which suggests
 that this experiment may already provide
  a hint about how the partial restoration of chiral symmetry
 manifest itself at finite density. 

In fact, a state of the art calculation based on the 
conventional many-body theoretical approach without
 incorporating the effect of the vacuum change 
was performed\cite{convention}; unfortunately, they all failed in 
reproducing the sufficient enhancement in the near-threshold
 region consistently with the other energy region.
Once the effect of partial chiral restoration  in nuclei
 is incorporated to the conventional approach\cite{chanf}, 
as suggested in \cite{HKS}, the agreement of the theory and experiment
 was remarkable. This is encouraging.

To confirm the threshold enhancement, first of all,
 more experimental work should be done.
Measurement of  2$\pi^0$ and 
$2\gamma$ in experiments with hadron/photon beams off
 the  heavy nuclear targets should be done,
 which is free from the $\rho$ meson meson background
  inherent in the $\pi^+\pi^-$ measurement.
Such an experiment was in fact performed by the Crystal  Ball(CB)
\cite{cb} 
collaboration
at BNL: They claimed that there is no threshold enhancement
that was seen in the CHAOS experiment. A reexamination of the
data by the CB group  has been done by the CHAOS group\cite{reexam},
and emphasized  the importance of  the combined ratio
\beq
C^{\bf A}_{\pi\pi}(M_{\pi\pi})=
\frac{\sigma^{\rm A}(M_{\pi\pi})/\sigma_{\rm T}^{\rm A }}
{\sigma^{\rm N}(M_{\pi\pi})/\sigma_{\rm T}^{\rm N }},
\eeq
where $\sigma_{\rm T}^{\rm A}\,(\sigma_{\rm T}^{\rm A})$
is the measured total cross section of the $\pi2\pi$ process
 in nuclei (nucleon): This ratio yields the net effect of
nuclear matter on the interacting $(\pi\pi)_{I=J=0}$ system.
They have shown that the combined ratio
grows  near the $2m_{\pi}$ threshold consistently in the
two experiments, although the statistics in the CB data is poorer.

Measuring of 2 $\gamma$'s from the electro-magnetic decay of the 
$\sigma$ or $(\pi\pi)_{I=J=0}$ in nuclear matter
may be interesting because of the small final state
 interactions, although the branching ratio is small.
One needs also to fight with large 
 background of photons mainly coming from $\pi^0$s.
 Nevertheless,  if the enhancement is prominent,
 there is a chance to find the signal.  
 When $\sigma$ has a finite three momentum,
one can detect dileptons  through the scalar-vector 
mixing in matter: $\sigma \to \gamma^* \to e^+ e^-$. 
The inverse process can be also used to produce the $\sigma$
or $(\pi\pi)_{I=J=0}$ system by the electro-magnetic probes
owing to the scalar-vector mixing in the finite density system 
where the charge conjugation symmetry is violated.
Such an experiment has been planned and being performed
in SPRING8\cite{SPRING8}.
We remark that (d, $^3$He) or (d, $^3$He)
 reactions is also useful to explore the
 spectral enhancement because of the
large incident flux.
 as in the production of the deeply bound pionic atoms
and the possible production of  $\eta$- or
$\omega$- mesic nuclei\cite{HHG}.
The incident kinetic energy $E$ of the  deuteron in the laboratory
system is  estimated to be  $1.1 {\rm GeV} < E < 10$ GeV, 
 to cover the spectral function 
 in the range  $2m_{\pi} < \omega < 750$ MeV.
A theoretical evaluation of the
feasibility of such experiments is now in progress\cite{hiren}.

\section{Other possible evidence of partial chiral restoration
 in nuclear matter}

It is interesting that there are other possible 
experimental evidences for partial chiral restoration in 
nuclear matter than the chiral fluctuations in the
sigma meson channel discussed so far.
The spectral function deduced from the lepton pairs from the
heavy ion collisions shows a softening, which might be an
evidence for the partial chiral restoration in nuclear
medium\cite{ceres}: Pisarski\cite{pisarski}
 was the first who suggested that
a decrease of the rho meson mass may be a signature of the 
chiral restoration in hot hadronic matter.
Brown and Rho\cite{br91} conjectured also the decrease of the vector meson
masses in association with the chiral restoration on the basis 
of a scaling argument (the so called Brown-Rho scaling).
Hatsuda and Lee\cite{hkl} discussed the vector meson properties using the
QCD sum rules.
A KEK experiment also shows the softening of the spectral
function in the $\rho/\omega$ channel in heavy nuclei such as
 Gold\cite{ozawa}.  The deeply bound pionic atom has proved 
to be a good probe of the properties of the hadronic interaction
 deep inside of heavy nuclei.  Yamazaki\cite{itahashi} suggested that
the anomalous energy shift of the pionic atoms (pionic nuclei)
owing to the strong interaction could be attributed to the
 decrease of the effective pion decay constant 
$f^{\ast}_{\pi}(\rho)$ at finite density $\rho$ which 
may imply that the chiral symmetry is partially restored deep
inside of nuclei.

\section{ Summary and concluding remarks}

\begin{enumerate}
\item The hadron spectroscopy must be a condensed matter
physics of the QCD vacuum, because  hadrons are elementary
excitations on top of the nontrivial QCD vacuum. 
\item The $\sigma$ meson as 
the quantum fluctuation of the
 order parameter of the chiral transition
 may account for various  phenomena in hadron physics
which otherwise remain mysterious.
\item There have been accumulation of 
{\em experimental evidence of the $\sigma$ pole} 
in the pi-pi scattering matrix.
Here it has been noticed that
the chiral symmetry, analyticity and crossing symmetry are 
all important.
\item
Partial restoration of chiral symmetry in hot and dense medium
 leads to a peculiar 
{\em enhancement in the spectral function
 in the $\sigma$ channel near the $2m_{\pi}$ threshold}.
\item The  enhancement is obtained both in the
linear and nonlinear realization of chiral symmetry
{\em provided that the
 possible reduction of the quark condensate or $f_{\pi}$
 is taken into account.}
\item Such an enhancement has been observed 
 in the reaction
A($\pi^{+}$, $(\pi^{+}\pi^{-})_{I=J=0}$)A' 
by CHAOS group, which  might possibly be
 an experimental
evidence of the partial restoration 
of chiral symmetry in heavy nuclei.
\item It seems that there is no serious contradiction between
 the CHAOS data $\pi^{+}\pi^{-}$ and the Crystal Ball data
on $2\pi^0$. 
\item Further theoretical and experimental works are needed
 to confirm that chiral symmetry is partially restored in heavy
nuclei.
\item There are other possible experimental evidences which
show a partial restoration in dense nuclear matter.
\end{enumerate}

{\bf Acknowledgments}
I thank the organizers of this
 symposium for inviting me to the symposium.
Most of this  report is based on the works done 
in collaboration with T. Hatsuda, D. Jido and H. Shimizu,
to whom I am grateful.
This work is partially supported  by the Grants-in-Aid of
the Japanese Ministry of Education, Science and Culture
(No. 12640263 and 12640296).

\end{document}